\documentclass[12pt]{article}
\usepackage{epsfig,latexsym}
\date{\today}
\def\bm{\boldmath}
\def\be{\begin{equation}}
\def\ee{\end{equation}}
\def\bear{\be\begin{array}}
\def\bea{\begin{eqnarray}}
\def\eea{\end{eqnarray}}

\def\lsi{\raise0.3ex\hbox{$<$\kern-0.75em\raise-1.1ex\hbox{$\sim$}}}
\def\gsi{\raise0.3ex\hbox{$>$\kern-0.75em\raise-1.1ex\hbox{$\sim$}}}
\def\lsim{\mathop{\lsi}}
\def\gsim{\mathop{\gsi}}

 \def\Pom{{\bf I\!P}}

\def\Qb{\overline{Q}}

\newsavebox{\fmbox}
    
\title {\bf \bm
The hard scale in the exclusive $\rho$ meson production in diffractive DIS}

 \author{I.P.~Ivanov\thanks{E-mail: i.ivanov@fz-juelich.de}\\
 \makebox[8cm][l]{\normalsize Institute of Mathematics, Novosibirsk, Russia}\\
 \makebox[8cm][l]{\normalsize IKP, Forschungszentrum J\"ulich, Germany}\\
 }

\begin{document}

\maketitle

 \vspace{-9cm}
 \makebox[\textwidth][r]{\large\bf FZJ-IKP(Th)-2003/XX}
 \vspace{7.5cm}

\abstract{We re-examine the issue of the pQCD factorization scale
in the exclusive $\rho$ production in diffractive DIS from the 
$k_t$-factorization point of view. We find that this scale
differs significantly from, and it possesses much flatter
$Q^2$ behavior than widely used value $(Q^2 + m_\rho^2)/4$. 
With these results in mind, we discuss
the $Q^2$ shape of the $\rho$ meson production cross section.
We introduce rescaled cross sections, which might provide further insight
into the dynamics of $\rho$ production. We also comment
on the recent ZEUS observation of energy-independent ratio
$\sigma(\gamma^{*} p \to \rho p) / \sigma_{tot}(\gamma^{*}p)$.}

\section{Introduction}

The exclusive production of vector mesons in diffractive DIS
$$
\gamma^{(*)} p \to Vp
$$
turned out to be an ideal testing ground \cite{VMfirst,NNNVM,NNZ94} 
of many predictions of the famous color-dipole approach, 
\cite{dipole1,dipole2,CTreviews}.
Within this formalism, the basic quantity is the
cross section of the color dipole interaction with the target proton
$\sigma_{dip}(x,\mbox{r})$, which can be approximately
related to the conventional gluon density $G(x_g,\Qb^2)$
inside the proton, \cite{dipole1}. The hard scale $\Qb^2$,
at which the gluon density should be taken, is set both by the 
virtuality and the mass of the vector meson, and in the case of 
heavy quankonium is approximately equal to ${1 \over 4}(Q^2 + m_V^2)$.
The latter result can be also obtained in the direct
DGLAP-inspired calculations of the relevant Feynman diagrams,
see, for example, the calculations of the exclusive
production of $J/\psi$ mesons in diffractive DIS \cite{ryskin}.

Although the identification of the pQCD factoriozation scale
with ${1 \over 4}(Q^2 + m_V^2)$ is valid only in the case of non-relativistic
vector mesons, there is a hope that the same relation
will hold for the light vector mesons and for large virtualities
as well. This hope leads to the prediction of a remarkable universality: 
the cross section of light and heavy mesons
will exhibit the same $Q^2$ behavior, if 
plotted against $Q^2 + m_V^2$ rather than $Q^2$ alone.

On the experimental side, the first data showed
that cross sections of $\rho$, $\omega$, $\phi$ and $J/\psi$
mesons, taken at equal values of $Q^2 + m_V^2$ and corrected
by corresponding $SU(4)$ factors, were indeed very close to each other,
\cite{clerbaux}. It is interesting to note that a similar
dependences on $Q^2 + m_V^2$ were observed not only in the magnitudes 
of the cross sections of different vector meson production, 
but also in the values of the energy growth exponent,
which is related to the Pomeron intercept, as well as
in the slopes of the forward diffraction cone.

However, with the advent of new, more accurate data of $\rho$ and $J/\psi$
production, it became clear that this scaling was only approximate.
Recently, ZEUS concluded \cite{levyDIS2002}
that the $J/\psi$ production cross section is typically 40\% higher
that the cross sections of the light vector meson production.
This difference is especially obvious on the plots of $\sigma_L$ and $\sigma_T$
separately, \cite{clerbaux}.

A natural question arises as to what extent the universality
of the $\Qb^2 = (Q^2 + m_V^2)/4$ as the relevant hard scale is valid.
Since there is almost no room for doubt in the case of $J/\psi$
production, the question can be formulated as {\em what is the
relevant pQCD factorization scale of the high $Q^2$ production of $\rho$ mesons}
\footnote{To be definite, we will deal with $\rho$ mesons, but the general
conclusions are obviously valid for all light vector mesons.}.

The answer to this question was, in fact, given already in \cite{NNZ94}.
Within the color dipole formalism, the pQCD factorization scales for the 
longitudinally and transverse produced $\rho$ mesons were shown to be
$\Qb^2(\rho_L) \approx 0.15\cdot (Q^2 + m_V^2)$ and 
$\Qb^2(\rho_T) \approx (0.07\div 0.1)\cdot(Q^2 + m_V^2)$.
This is notably smaller than $(Q^2 + m_V^2)/4$, and suggests
that even the highest $Q^2$ experimental points in $\rho$ production
correspond to, at most, semiperturbative values of $\Qb^2(\rho)$. 

In fact, the exact value of the pQCD factorization scale
can be affected by the shape of the unintegrated gluon distribution.
Unfortunately, when \cite{NNZ94} appeared, 
no numerically reliable parametrizations
of dipole cross section or of the unintegrated gluon structure
function were available, and one was bound to a semiquantitative
guess. The situation changed two years ago,
when numerially accurate, simple, and ready-to-use parametrizations
of the unintegrated gluon structure function
${\cal F}(x_g,\vec\kappa)$ were obtained from the analysis
of proton structure function $F_{2p}$, \cite{IN2001}.
These parametrizations were devised for $x_{Bj} < 0.01$ and
for the entire domain of relevant $Q^2$ values.
They were put in the basis of the $k_t$-factorization calculations
of both light and heavy vector meson production cross sections
and yielded rather good description 
of the avaliable data, \cite{INS2003,VMtalks,disser}.
These fits now allow for a quantitative reanalysis
of the hard scale in the $\rho$ meson production.
This is performed in the present paper.\\

Numerically accurate understanding of the hard scale
in the $\rho$ production is demanded in many applications.
For example, a need for the hard scale arises 
when one attempts to understand the $Q^2$ behavior
of the $\rho$-meson electroproduction. The early data
could be successfully parametrized by a simple law
$$
\sigma(\gamma^* p\to \rho p) \propto {1 \over (Q^2 + m_\rho^2)^n}
$$
with $n = 2.32 \pm 0.10$ (ZEUS, \cite{ZEUSrho}) or 
$n = 2.24 \pm 0.09$ (H1, \cite{H1rho}). However, 
with the advent of high precision data in a much broader
$Q^2$ region, it became clear that such powerlike fits 
have very limited applicability domain. A natural question has been 
raised as what would be the most insightful and physically motivated
fit to the $Q^2$ behavior of $\sigma(\gamma^* p \to \rho p)$.

In virtually any description of the exclusive $\rho$ meson production,
one has to deal with the gluon content of the proton.
Clearly, in order to experimenally study the underlying mechanism
of the vector meson production, one might want to get rid of the ``trivial''
source of the $Q^2$ behavior that arises due to the gluon density.
A simple way to do this would be to divide the cross sections
by the conventional gluon density squared and study
the $Q^2$ behavior of the result. However, this procedure 
requires knowledge of the hard scale, at which the gluon 
density should be calculated.

Another issue that demands the understanding of the hard scale
in $\rho$ production is related to the energy behavior
of the ratio
\be
r_\rho = {\sigma(\gamma^{(*)} p \to \rho p) 
\over \sigma_{tot}(\gamma^{(*)}p)},
\ee
The pQCD improved Regge model predicts that these cross sections
rise with energy as
\be
{\sigma(\gamma^{(*)} p \to \rho p) \propto (W^2)^{2\Delta_\Pom}\,;\quad
 \sigma_{tot}(\gamma^{(*)}p)} \propto (W^2)^{\Delta_\Pom}\,,\label{test1}
\ee
where $\Delta_\Pom \equiv \alpha_\Pom -1$ is the Pomeron intercept,
so that the ratio is predicted to rise as
\be
r_\rho \propto (W^2)^{\Delta_\Pom}\,.\label{test2}
\ee
In last years, a large amount of data became available
on the $\rho$ meson photo- and electroproduction in a broad
range of the total photon-proton energy $W$ and the photon 
virtuality $Q^2$, \cite{ZEUSrho,H1rho}.
This enabled the ZEUS collaboration to study the energy behavior
of $r_\rho$ at several values of $Q^2$.
The results, \cite{levyDIS2002}, posed an apparent challenge:
throughout the whole $Q^2$ range, 
from $Q^2 = 0$ up to $Q^2 = 27$ GeV$^2$, the ratio $r_\rho$
was found to be energy independent.
The conclusion of \cite{levyDIS2002} was that neither Regge model,
nor pQCD approach can explain this constancy.
 
However, the Pomeron intercept is known to significantly depend on the 
hard scale involved in the interaction, 
see experimental data \cite{ZEUSrho,H1rho,ZEUSF2,H1F2} and
results of the phenomenological analysis of \cite{IN2001}. 
Thus, when testing prediction (\ref{test2}), one must make sure that
both intercepts in (\ref{test1}) are taken at the same hard scale.

In \cite{levyDIS2002} these scales were identified with $Q^2$ for
the total virtual photoabsorption cross section and with 
$\Qb^2_\rho = \Qb^2 \equiv (Q^2 + m_\rho^2)/4$ for the $\rho$ meson 
production cross section, so that the experimental results
were given for the quantity:
$$
r_\rho = {\sigma(\gamma^{(*)} p \to \rho p)(W^2,Q^2) 
\over \sigma_{tot}(\gamma^{(*)}p) (W^2,\Qb^2)}.
$$
Note that here the $\rho$ meson production cross section
and the total virtual photoabsorption cross section
are taken at different virtualities. 

As we argue in this paper,
the true scale of the $\rho$ production can noticeably deviate
from $(Q^2 + m_\rho^2)/4$, especially at very small and very large $Q^2$.
Thus, it appears that the mismatch of the scales can 
be at least one of the sources of the discrepancy observed.\\

The structure of the paper is the following.
In Section 2 we briefly review the results of the
$k_t$-factorization approach to the exclusive production 
of vector mesons in diffractive DIS. For future guidance, we
also show how these results simplify in the case of 
heavy vector mesons. In Section 3 we study which gluons
are relevant for the interaction, and settle
the factorization scales $\Qb^2_L$ and $\Qb^2_T$ for longitudinal
and transverse amplitudes for any $Q^2$. Section 4 serves as an application
of the results obtained to the study of the $Q^2$ behavior
of the $\rho$ meson electroproduction cross section.
In this Section we also propose new quantites, the rescaled cross sections,
which might provide additional insight into the mechanism
of the interaction. Finally, Section 5 is devoted to the
discussion of the results and to the conslusions.

\section{Exclusive production of vector mesons}

Within the familiar color dipole formalism \cite{dipole1,dipole2}, 
the production of a vector meson
is viewed as a three-step process. First, at distances
of the order of the coherence length 
$\ell_c \sim 1/(m_N x_{Bj})$ (where $x_{Bj}$ is Bjorken $x$
and $m_N$ is the nucleon mass) upstream the target, 
the incident photon splits into a $q\bar q$ pair, a color dipole. 
Then, interaction between the color dipole and the target takes place.
This $q\bar q$ pair receives longitudinal momentum transfer, 
so that its invariant mass squared turns positive and close
to the mass of a corresponding vector meson. Then, at large
formation distances, this $q\bar q$ pair projects onto hadronic 
final states.

At high energies, the typical longitudinaly size of the interaction
region, of the order of radius of the proton, is much smaller
than both the coherence length and the formation length.
This allows one to treat the color dipole frozen during the
interaction, and the basic quantity that appears in the approach 
is the diagonal color dipole cross section $\sigma_{dip}(x,r)$, which is
related to the gluon density of the target proton. 
The analysis of \cite{NNZ94} yielded the result that 
the pQCD factorization scales for the 
longitudinally and transverse produced $\rho$ mesons were
$\Qb^2(\rho_L) \approx 0.15\cdot (Q^2 + m_V^2)$ and 
$\Qb^2(\rho_T) \approx (0.07\div 0.1)\cdot(Q^2 + m_V^2)$.

In this paper we reexamine this issue within the $k_t$-factorization
approach, which is closely related to the color dipole formalism.
As said above, our analysis is based on the parametrizations 
of the unintegrated gluon structure
function that were given in \cite{IN2001}.

\subsection{The basic amplitude}

\begin{figure}[!htb]
   \centering
   \epsfig{file=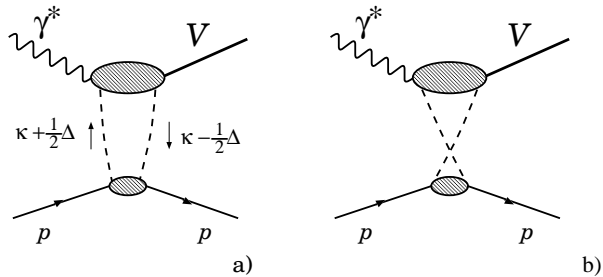,width=80mm}
   \caption{The QCD--inspired diagrams for $\gamma^* p \to Vp$ process
with two gluon $t$--channel. Only Diagr.(a) contributes to the
imaginary part of amplitudes.}
   \label{diagram1}
\end{figure}

\begin{figure}[!htb]
   \centering
   \epsfig{file=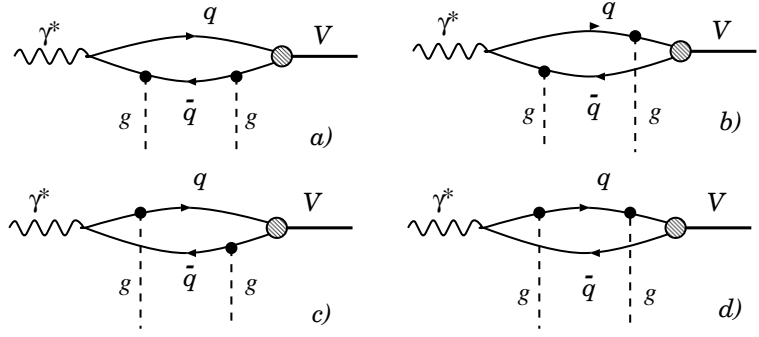,width=100mm}
   \caption{The content of the upper blob in Fig.\ref{diagram1}a in the
pQCD approach.}
   \label{diagram2}
\end{figure}

Within the $k_t$-factorization approach, the imaginary part
of the amplitude of the vector meson production is given by 
discontinuity of the diagram shown in Fig.~\ref{diagram1}a.
The upper blob should be understood as sum of four diagrams, 
shown in Fig.~\ref{diagram2}.
We denote the quark and gluon loop transverse momenta and the momentum transfer 
by $\vec k$, $\vec \kappa$, and $\vec\Delta$, respectively. 
The fraction of the photon lightcone momentum carried by quark is denoted
by $z$, while the fractions of the proton light cone momentum carried
by the two gluons are $x_1$ and $x_2$. 
With this notation, the imaginary part of the amplitude
can be written in a compact form:
\be
Im {\cal A}= s{c_{V}\sqrt{4\pi\alpha_{em}}\over 4\pi^{2}}
\int {d^{2} \vec \kappa \over \vec\kappa^{4}}\alpha_{S}(q^2)
{\cal{F}}\left(x_1,x_2,\vec\kappa_1,\vec\kappa_2\right)
\int {dzd^2 \vec k \over z(1-z)}  \psi^*_V(z,\vec k)
\cdot I(\lambda_V,\lambda_\gamma)\, .
\label{f1}
\ee
where $\vec\kappa_{1,2} = \vec \kappa \pm {1 \over 2}\vec \Delta$.
The helicity-dependent integrands $I(\lambda_V,\lambda_\gamma)$ 
have form
\begin{eqnarray}
I^S(L,L) &=& 4 QM z^2 (1-z)^2
\left[ 1 + { (1-2z)^2\over 4z(1-z)} 
{2m \over M+2m}\right] \Phi_2\,;\nonumber\\[1mm]
I^S(T,T) &=& (\vec{e}\vec{V}^*)[m^2\Phi_2 + (\vec{k}\vec{\Phi}_1)] 
+ (1-2z)^2(\vec{k}\vec{V}^*)(\vec{e}\vec{\Phi}_1){M \over M+2m}\nonumber\\
&&- (\vec{e}\vec{k})(\vec{V}^*\vec{\Phi}_1) + 
{2m \over M+2m}(\vec{k}\vec{e})(\vec{k}\vec{V}^*)\Phi_2\,;\nonumber\\[1mm]
I^S(L,T) &=& 2Mz(1-z)(1-2z)(\vec{e}\vec{\Phi}_1)
\left[ 1 + { (1-2z)^2\over 4z(1-z)} {2m \over M+2m}\right]
- {Mm\over M+2m}(1-2z)(\vec{e}\vec{k})\Phi_2\,;\nonumber\\[1mm]
I^S(T,L) &=& -2Qz(1-z)(1-2z)(\vec{V}^*\vec{k}){M \over M+2m}\Phi_2\,.\label{f8}
\end{eqnarray}
where
\begin{eqnarray}
\Phi_2& =& -{1 \over (\vec{r}+\vec\kappa)^2 + \varepsilon^2} -{1 \over
(\vec{r}-\vec\kappa)^2 + \varepsilon^2} + {1 \over (\vec{r} + \vec\Delta/2)^2 +
\varepsilon^2} + {1 \over (\vec{r} - \vec\Delta/2)^2 + 
\varepsilon^2}\nonumber\,;\\[5mm]
\vec{\Phi}_1 &=& -{\vec{r} + \vec\kappa \over (\vec{r}+\vec\kappa)^2 + \varepsilon^2}
-{\vec{r} - \vec\kappa \over (\vec{r}-\vec\kappa)^2 + \varepsilon^2}
+ {\vec{r} + \vec\Delta/2 \over (\vec{r} + \vec\Delta/2)^2 + \varepsilon^2}
+ {\vec{r} - \vec\Delta/2 \over (\vec{r} - \vec\Delta/2)^2 + 
\varepsilon^2}\nonumber\,,
\end{eqnarray}
with $\vec r \equiv \vec k - (1-2z)\vec\Delta/2$ and 
$\varepsilon^2 = z(1-z)Q^2 + m_q^2$. Finally, the strong coupling
constant is taken at 
$q^2 \equiv \mbox{ max}[\varepsilon^2 + \vec k^2, \vec\kappa^2]$.

In the absence of $\vec\Delta-\vec\kappa$ correlations,
and for a very asymmetric gluon pair, the off-forward gluon structure function 
${\cal{F}}\left(x_1,x_2,\vec\kappa_1,\vec\kappa_2\right)$ 
that enters (\ref{f1}) can be approximately related to the forward 
gluon density ${\cal F}(x_g,\vec \kappa)$ via
$$
{\cal F}(x_1,x_2\ll x_1,\vec\kappa_1,\vec\kappa_2) 
\approx {\cal F}(x_g,\vec\kappa) \cdot 
\exp\left(- {1 \over 2}b_{3\Pom}\vec\Delta^2\right)\,.
$$
Here $x_g \approx 0.41 x_1$; the coefficient $0.41$ 
is just a convenient representation
of the off-forward to forward gluon structure function relation
found in \cite{shuvaev}.
Numerical parametrizations of the forward unintegrated gluon density 
 ${\cal F}(x_g,\vec \kappa)$ for any practical values of $x_g$ and 
$\vec \kappa^2$ can be found in \cite{IN2001}.
The slope $b_{3\Pom}$ can be experimentally accessed in the
high-mass elastic diffraction.

The vector meson wave function $\psi_V(z,\vec k)$ describes 
the projection of the $q\bar q$ pair onto the physical vector meson.
It is normalized so that the forward value of the vector meson
formfactor is unity, and the free parameters are chosen 
to reproduce the experimentally observed value of the
vector meson electronic decay width $\Gamma(V \to e^+e^-)$.
In what concerns the shape of the radial wave function,
we followed a pragmatic strategy. Namely, we took a simple
Ansatz for the wave function, namely, the oscillator type wave function
and performed all calculations with it. In order
to control the level of uncertainty,
introduced by the particular choice of the wave function,
we redid calculations with another wave function Ansatz, namely,
the Coulomb wave function, and compared the results.
Since these two wave functions represent the two extremes (very compact
and very broad wave functions that still lead to the same value of 
the electronic decay width), this difference gives
a reliable estimate of the uncertainty.
It is given typically by factor of 1.5 for magnitudes
of the cross sections, while in the observables that involve
ratios of the cross sections (including slopes and intercepts) 
this uncertainty is reduced.
Details can be found in \cite{disser,INS2003}.

Note also that when deriving (\ref{f8}), we treated the vector mesons
as $1S$ wave states and used the pure $S$-wave spinorial structure 
${\cal S}^\mu$ instead of $\gamma^\mu$, \cite{IN99}.

\subsection{The heavy meson limit and the scaling property}

It is instructive to look at the production amplitude (\ref{f1}) 
in the heavy quarkonium limit. In this case, the vector mesons
is essentially non-relativistic, $z \to 1/2\,, \vec k^2 \ll \vec m_q^2$, 
and
$$
\varepsilon^2 = z(1-z)Q^2 + m_q^2 \approx {1 \over 4}(Q^2 + m_V^2)
\equiv \Qb^2\,.
$$
Even in the photoproduction limit,
the mass of the meson sets the hard energy scale of the interaction,
and one should expect the DGLAP-inspired leading log $\Qb^2$ approximation
to be a viable approach to the problem.
The DGLAP approximation of the heavy vector meson production
implies the following ordering
$$
\vec k^2 \ll \vec\kappa^2 \ll \Qb^2\,,
$$
which leads to simplifications in the production amplitude.
Consider the forward scattering. In this case only helicity conserving
amplitudes survive, and one obtains
$$
I(L,L) = \vec\kappa^2\,{8 Q m_V \over (Q^2 + m_V^2)^2}\,;\quad
I(T,T) = \vec\kappa^2\,{8 m_V^2 \over (Q^2 + m_V^2)^2}\,.
$$
We reiterate that this simplification works while $\vec\kappa^2$ is sufficiently
small, namely, $\vec\kappa^2 \ll \Qb^2$.
In the spirit of the DGLAP approach, the integrands $I(\lambda_V,\lambda_\gamma)$
do not depend on $\vec\kappa^2$, aside from the overall $\vec\kappa^2$
factor, whose origin is the color neutrality of the proton.
Then, integration of $\vec\kappa^2$ immediately leads to the
conventional gluon density:
\bea
\int_0^\infty {d\kappa^2 \over \kappa^4} \alpha_s(\vec\kappa^2) 
{\cal F}(x_{g},\vec\kappa^2) I(T,T) \quad \to\quad
{8 m_V^2 \over (Q^2 + m_V^2)^2}
\int_0^{\Qb^2} {d\kappa^2 \over \kappa^2} \alpha_s(\vec\kappa^2) 
{\cal F}(x_{g},\vec\kappa^2)\nonumber\\ \approx
{8 m_V^2 \over (Q^2 + m_V^2)^2}\cdot \alpha_s(\Qb^2)\cdot 
G(x_{g},\Qb^2)\,.\nonumber
\eea
The integration of the wave function can be related to the electronic decay
widths of the vector meson. The result for the forward differential
cross section reads, \cite{ryskin}:
\be
{d\sigma \over d|t|}\Bigg|_{t = 0} = { \pi^3 \over 12 \alpha_{em}}
m_V \,\Gamma(V\to e^+e^-)\, {\left[\alpha_s(\Qb^2)\cdot 
G(x_{g},\Qb^2)\right]^2 \over \Qb^6}\,.\label{VMheavy}
\ee
Note that in this limit the difference between the pure $S$-wave
vector meson and the vector meson with $\gamma^\mu$ coupling
vanishes.

Eq.(\ref{VMheavy}) identifies the hard scale that 
defined the gluon content of the proton in such a reaction
with the quantity $\Qb^2 = (Q^2 + m_\rho^2)/4$. Although
this derivation is valid only for very heavy vector mesons,
it is often hoped that for the light mesons the hard scale will 
have the same form. We show below that this is not true.

This result, although an approximation, suggests the scaling property
of the vector mesons production: the $Q^2$ behavior
of all vector mesons cross sections should be proportional to each other
if taken at equal $Q^2 + m_V^2$.
Note that (\ref{VMheavy}) {\em does not} suggest that
the {\em absolute values} of the cross sections should be equal,
even when corrected by the relevant flavor content factors $c_V^2$,
simply due to the fact that this approach does not
predict universality of $m_V \cdot \Gamma(V\to e^+e^-)$.

\section{The scale in exclusive $\rho$ production}

In this section we perform an analysis of the amplitude (\ref{f1}),
taken for simplicity at $\vec\Delta = 0$, and find
what values of the transverse gluon  momenta are essential 
in the interaction. These values will define the hard scale 
of the conventional gluon density. Again, this analysis is essentially 
similar to what was done in \cite{NNZ94}.

\subsection{Weight functions: mapping the glue}

We start with a qualitative analysis 
of the production of longitudinal $\rho$  mesons at high $Q^2$.
Let us for a moment neglect the scaling violations and
the other sources of marginal $\vec k^2$ behavior, and 
rewrite the amplitude (\ref{f1}) in the following simplified form:
$$
Im A_{L\to L} \propto  \int{d\vec\kappa^2 \over \vec\kappa^2}
{\cal F}(x_g,\vec\kappa) \cdot W_L(\vec\kappa^2)
\,;\quad W_L \approx {1 \over \vec\kappa^2}
\int_{0}^{1} {dz\,z(1-z)} \int d^2 \vec k \psi(z,\vec k)\, \Phi_2
$$ 
with
$$
\Phi_2 = -{1 \over (\vec{k}+\vec\kappa)^2 + \varepsilon^2} -{1 \over
(\vec{k}-\vec\kappa)^2 +  \varepsilon^2} + {2 \over \vec{k}^2 +
 \varepsilon^2}\,.
$$
At large $Q^2$, the most of the $\vec\kappa^2$ integration
comes from the logarithmic region with 
$\vec k^2 \ll \vec\kappa^2 \ll \varepsilon^2$.
This leads to simplifications:
\be
W_L(\vec\kappa^2) \approx 
\int_{0}^{1} {dz\,z(1-z)} \int d^2 \vec k \psi(z,\vec k)\cdot 
{2 \over \varepsilon^2 (\varepsilon^2 + \vec\kappa^2)}\, \label{wlapprox}
\ee
If we denote by  $\langle \varepsilon^2\rangle$
the typical values of $\varepsilon^2 = z(1-z)Q^2 + m_q^2$  
that dominate the integral (\ref{wlapprox}), then the 
weight factor $W_L(\vec\kappa^2)$ stays almost constant at 
$\vec\kappa^2 \ll \langle \varepsilon^2\rangle$, and
quickly decreases for $\vec\kappa^2 >  \langle \varepsilon^2\rangle$.
It should have a form of ``smoothed step function",
and effectively cuts off from above the $\vec \kappa^2$ region 
essential for the interaction. 
If we further approximate it by the exact step 
function $\theta(\vec\kappa^2 - \Qb_L^2)$ with properly defined
cut-off scale $\Qb^2_L$, then the integration (\ref{wlapprox})
can be done exactly:
\be
\int_0^\infty {d\vec\kappa^2 \over \vec\kappa^2}
{\cal F}(x_{g},\vec\kappa) W_L(\vec\kappa^2)
\Rightarrow W_L(0)\cdot \int_0^{\Qb_L^2} {d\vec\kappa^2 \over \vec\kappa^2}
{\cal F}(x_{g},\vec\kappa) = 
W_L(0) \cdot G(x_{g},\Qb_L^2)\,.\label{int3}
\ee
This transition is precisely what is implied in the DGLAP approach,
recall derivation of (\ref{VMheavy}).
Thus, a procedure that would yield the values of the relevant hard scale,
at which the conventional gluon density should be taken,
is the following: take the weight function $W_L(\vec\kappa^2)$ and
find the value $\Qb^2_L$ that would lead to the most justified
replacement (\ref{int3}).

Obviously, the scale $\Qb^2_L$ should be close to
$\langle \varepsilon^2 + \vec k^2\rangle$.
Due to the broad wave function of the $\rho$, 
the typical values of $z(1-z)$ will be less
than $1/4$. As a result, the scale of the gluon density
in (\ref{int3}) should be noticeably softer than $(Q^2 + m^2_\rho)/4$.

In the case of the production of the transversely polarized 
$\rho$ mesons, the contribution of the very asymmetric $q\bar q$ pairs
is enhanced. This leads to typical values of $z$ still smaller, 
and the departure of the corresponding factorization scale $\Qb^2_T$ 
from $(Q^2 + m^2_\rho)/4$ becomes even more prominent.

\subsection{Numerical results}

\begin{figure}[!htb]
   \centering
   \epsfig{file=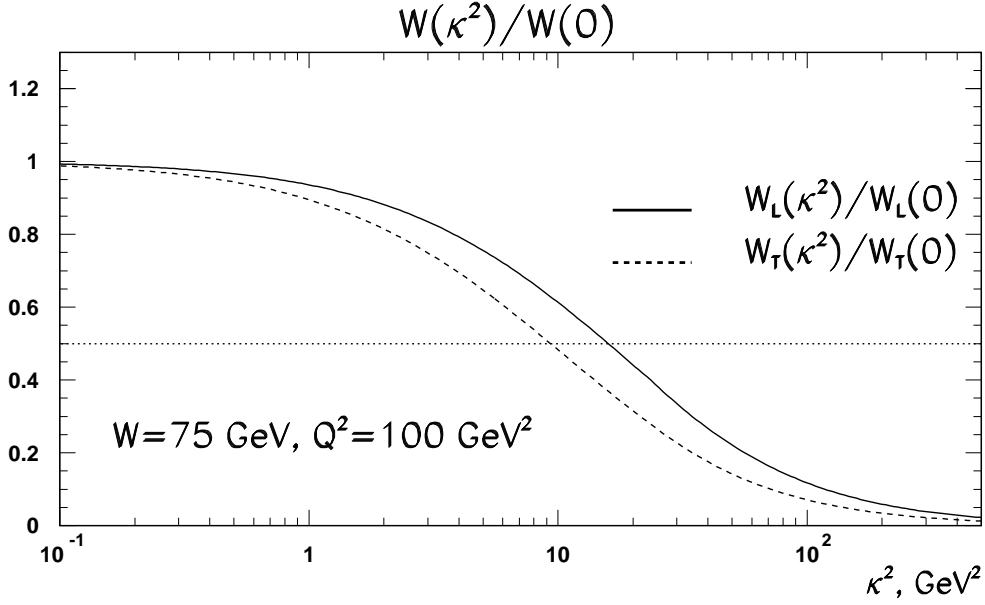,width=14cm}
   \caption{\em The normalized weight functions 
$W_L(\vec\kappa^2)/W_L(0)$ and $W_T(\vec\kappa^2)/W_T(0)$
calculated at $Q^2 = 100$ GeV$^2$. The $\vec\kappa^2$ values where
$W_L$ and $W_T$ reach $1/2$ are noticeably softer than $(Q^2+m_\rho^2)/4$.}
   \label{w-rho}
\end{figure}

The above expectations are corroborated by the numerical
analysis. The functions $W_L(\vec\kappa^2)$ and $W_T(\vec\kappa^2)$ 
are defined now via
$$
{1 \over s}Im A_{L\to L} \equiv \int{d\vec\kappa^2 \over \vec\kappa^2}
{\cal F}(x_g,\vec\kappa) \cdot W_L(\vec\kappa^2)\,;\quad
{1 \over s}Im A_{T\to T} \equiv \int{d\vec\kappa^2 \over \vec\kappa^2}
{\cal F}(x_g,\vec\kappa) \cdot W_T(\vec\kappa^2)
$$
for all values of $Q^2$. Their behavior for $Q^2 = 100$ GeV$^2$ 
is illustrated in Fig.~\ref{w-rho}, where we show 
the ratios $W_L(\vec\kappa^2)/W_L(0)$ and $W_T(\vec\kappa^2)/W_T(0)$ 
as a functions of $\vec\kappa^2$.
One sees that these weight functions start decreasing at $\vec \kappa^2 \ll
Q^2$ and reach $1/2$ at $\vec\kappa^2 \approx 15$ GeV$^2$ 
and $\vec\kappa^2 \approx 10$ GeV$^2$,
respectively. This shows that the above mentioned effect of the broad 
wave function leads to significant softening of the relevant scale.

The factorization scales $\Qb^2_L$ and $\Qb^2_T$ can be defined,
in principle, in several ways. As a trial definition,
one can look at the $\vec\kappa^2$ points, 
where $W_i(\vec\kappa^2)$ reach ${1 \over 2}W_i(0)$.
Defined so, the longitudinal and transverse 
scales were found to be approximately equal to
${1 \over 6}(Q^2 + 2.0 \mbox{ GeV}^2)$ and 
${1 \over 11}(Q^2 + 2.6 \mbox{ GeV}^2)$, respectively.
The numbers ${1 \over 6}$ and ${1\over 11}$ are very close to
$0.15$ and $0.07 \div 0.1$  obtained in \cite{NNZ94}.

The gluon density, however, has itself significant $\vec\kappa^2$ 
dependence, \cite{IN2001}. Namely, in the region 
$\vec\kappa^2 \sim 1\div 10$ GeV$^2$ and very small $x_g$ ($x_g \lsim 10^{-3}$)
(which corresponds, at fixed $W = 75$ GeV, 
to values of $Q^2 \lsim 10$ GeV$^2$),
${\cal F}(x_g,\vec\kappa^2)$ is a strongly rising function 
of $\vec \kappa^2$. At larger $Q^2$, the effective $x_g$ grows,
and the unintegrated gluon density becomes flat.
Finally, at large enough $Q^2$ (for $W = 75$ GeV, this 
corresponds to $Q^2 \gsim 100$ GeV$^2$), the unintegrated gluon density
decreases with $\vec\kappa^2$ growth in the region 
$\vec\kappa^2 \sim 1\div 10$ GeV$^2$.
Therefore, the span of effectively contributive $\vec\kappa^2$
will extend to higher values of $\vec\kappa^2$ (at small $Q^2$) 
or reduce to smaller values of $\vec\kappa^2$ (at high $Q^2$). 

In order to take this into account, it is more useful 
to define the hard scales via the following implicit relations:
\be
{1 \over W_i(0)} \int_0^\infty {d\vec\kappa^2 \over \vec\kappa^2}
{\cal F}(x_{g},\vec\kappa) W_i(\vec\kappa^2)
\equiv G(x_{g},\Qb_i^2)\,;\quad i = L,\,T\label{int4}
\ee

\begin{figure}[!htb]
   \centering
   \epsfig{file=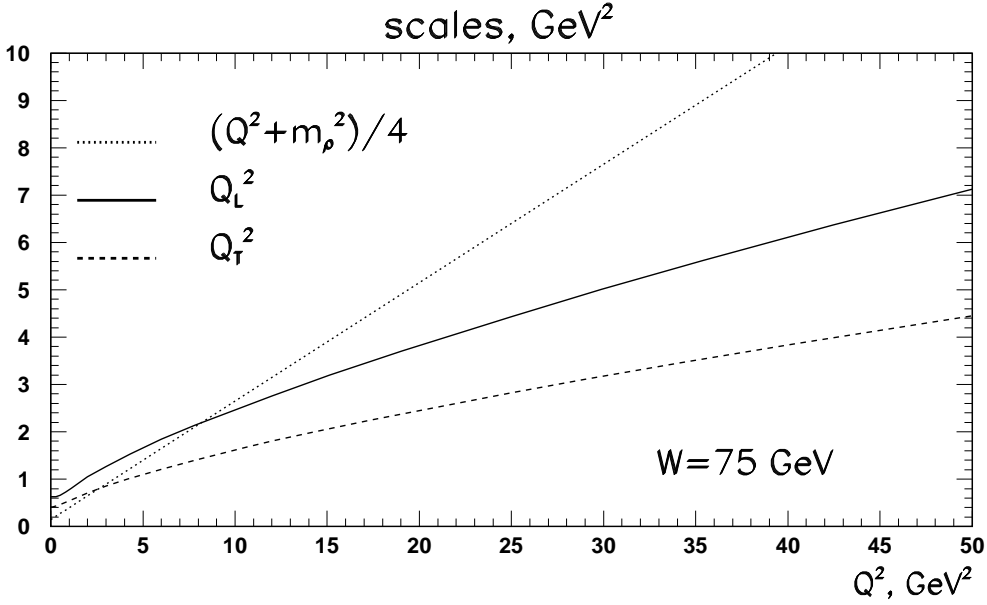,width=14cm}
   \caption{\em The $Q^2 \to \Qb^2_L$ and $Q^2 \to \Qb^2_T$ mapping 
in the $\rho$ production. The heavy meson analysis expectation
$(Q^2 + m_\rho^2)/4$ is also shown.}
   \label{scale}
\end{figure}

Fig.~\ref{scale} shows the values of $\Qb^2_L$ (solid line) and $\Qb^2_T$
(dashed line), defined according to (\ref{int4}), as functions of $Q^2$. 
These values start from $0.63$ GeV$^2$ and $0.4$ GeV$^2$, respectively,  
in the photoproduction limit, and slowly grow with $Q^2$ rise. 
At $Q^2 = 27$ GeV$^2$ (the highest $Q^2$ data point from H1 data
on $\rho$ production), these values are
only around 4.5 GeV$^2$ and 3 GeV$^2$, respectively.
This confirms the conclusion of \cite{NNZ94} that
even at largest $Q^2$ where data are available, we still
deal with semiperturbative situation.
It is interesting to note that a better fit to these curves
is given by a non-linear, rather than linear approximation:
\be
\Qb_L^2 \approx 1.5\cdot \Qb_T^2 \approx 0.45\cdot (Q^2 + 1.5)^{0.7}\,,
\label{qb2param}
\ee
where all quantities are expressed in GeV$^2$.

The same Figure shows also, by dotted line, the expectation $(Q^2 + m_\rho^2)/4$
inspired by the heavy meson analysis.
This expectation starts from $0.15$ GeV$^2$, which is noticeably smaller
than $\Qb^2_L(0)$ and $\Qb^2_T(0)$, and rises with $Q^2$ 
significantly faster than $\Qb^2_L$ and $\Qb^2_T$.

\subsection{The $Q^2$ dependence of the weight functions}

In the heavy meson approximation (\ref{VMheavy}), the scale 
$\Qb^2 = (Q^2 + m_V^2)/4$ defines not only the hard scale of the gluon density,
but also the absolute value of the cross section.
It is interesting to check whether such a universality holds
for the $\rho$ production.

This can be done by studying the $Q^2$ dependence of the $W_L(0)$ and
$W_T(0)$. Motivated by (\ref{VMheavy}), we introduce the ``effective twists"
$\varepsilon_L^2$ and $\varepsilon_T^2$ via the following relations: 
\be
W_L(0) = 0.153\; Q \cdot {\alpha_s(\varepsilon^2_L) \over (\varepsilon^2_L)^2}\,;\quad
W_T(0) = 0.153\; m_V \cdot {\alpha_s(\varepsilon^2_T) \over (\varepsilon^2_T)^2}\,.
\ee
Here, factor $0.153$ GeV $ = \sqrt{\Gamma(\rho \to e^+e^-) \pi^4 m_\rho 
\over 3\alpha_{em}}$ is the numerical value of all the residual
factors present in the definition of $W_i$.
This definition of the ``effective twists" $\varepsilon_L^2$ and $\varepsilon_T^2$
is such that in the heavy meson limit one would recover the familiar 
$\Qb^2 = 0.25\cdot(Q^2 + m_V^2)$. 
These ``effective twists" were found to be approximately equal to
\be
\varepsilon^2_L \approx 0.23\cdot (Q^2 + 1.1)\,;\quad
\varepsilon^2_T \approx 0.16\cdot (Q^2 + 1.3)\,.\label{vareps}
\ee
Again, all quantities are expressed here in GeV$^2$.
It is interesting to note that here, as well as in (\ref{qb2param}),
the virtuality $Q^2$ sums with 
$M^2 \sim 1\div 1.5$ GeV$^2$ instead of widely assumed $m_\rho = 0.6$ 
GeV$^2$. This is not surprising, 
since the mass of the $\rho$ meson does not have much relevance
to the interaction of the $q\bar q$ dipole with the proton.

\section{The $Q^2$ behavior of the $\rho$ production cross section}

Let us turn to an issue that is closely related to the above
discussion of the hard scales in the $\rho$ production, namely, 
to the $Q^2$ behavior of the exclusive $\rho$ meson 
production cross section.

As mentioned in the Introduction, the early data on $\rho$ mesons
were successfully parametrized (in the moderate and high-$Q^2$ region) 
by a simple law
$$
\sigma(\gamma^* p\to \rho p) \propto {1 \over (Q^2 + m_\rho^2)^n}
$$
with $n = 2.32 \pm 0.10$ (ZEUS, \cite{ZEUSrho}) or 
$n = 2.24 \pm 0.09$ (H1, \cite{H1rho}). However, subsequent data
made it clear that such power fits have very limited 
applicability. One thus can ask what would be 
the best way to parametrize this behavior.
Note that this is not merely a question of how to fit the data,
but rather a question of what are the main sources of $Q^2$ behavior
of the cross section.

Let us take the heavy meson limit result (\ref{VMheavy}) as
a starting point. It states that the cross section should quickly fall
with $Q^2$ due to powers of $Q^2 + m_V^2$, the fall
being just slightly tamed by the rise of the gluon density.
In order to check whether or not these are the only sources
of the $Q^2$ behavior of the $\rho$ meson cross section,
let us consider a {\em rescaled cross section}
\be
\Sigma(Q^2) = \sigma(\gamma^* p \to \rho p)\cdot 
{(\Qb^2)^3 \cdot b(Q^2)\over \left[G\left(x_g,\Qb^2 \right)
\cdot \alpha_s(\Qb^2)\right]^2}\,.\label{rescaled1}
\ee
Here, as before, $\Qb^2 \equiv (Q^2 + m_\rho^2)/4$.
Note again that the gluon density is taken here 
at constant energy ($W = 75 $ GeV), so that 
$x_g = 0.41 (Q^2 + m_\rho^2)/W^2$ also depends on $Q^2$.
From the heavy meson limit expression (\ref{VMheavy}) 
one might hope that $\Sigma(Q^2)$ is roughly independent
of $Q^2$, since all the sources of $Q^2$ dependence
are removed.

\begin{figure}[!htb]
   \centering
   \epsfig{file=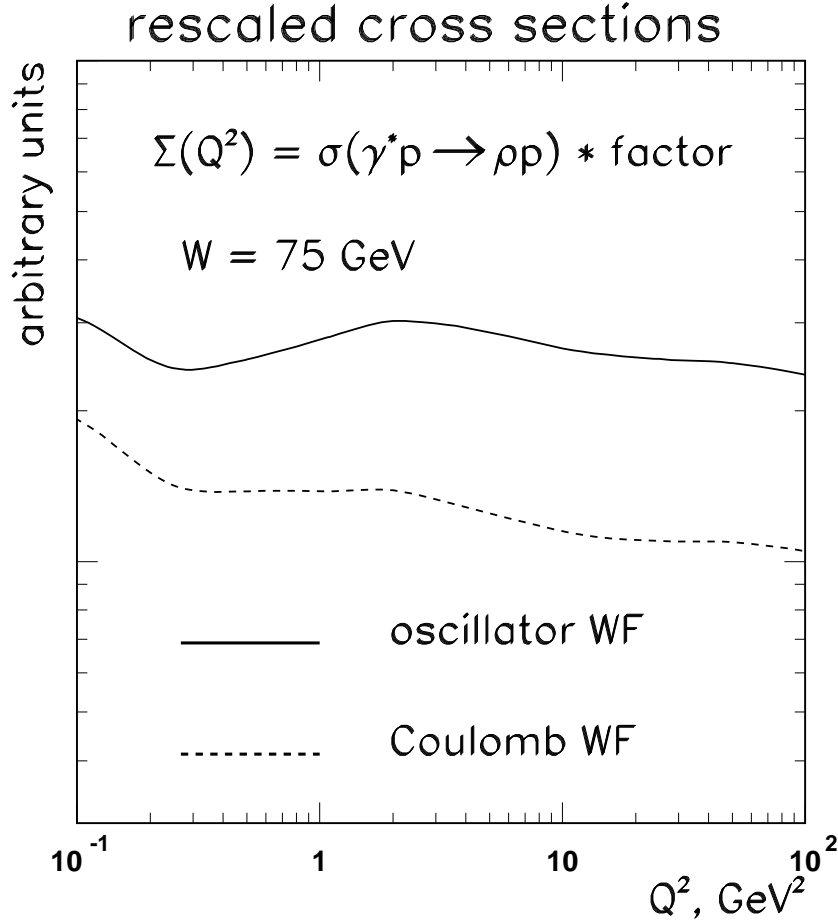,width=12cm}
   \caption{\em The rescaled cross section $\Sigma(Q^2)$ defined according
to (\ref{rescaled1}).}
   \label{figrescaled1}
\end{figure}

Fig.~\ref{figrescaled1} shows the $k_t$-factorization predictions 
for $\Sigma(Q^2)$ as function of $Q^2$.
One observes that this rescaled cross section has quite a flat shape,
from very low to very high $Q^2$,
regardless of the exact wave function used.

In principle, all the observables used in (\ref{rescaled1})
are accessible in experiment. The only delicate issue
would be the choice of the gluon density, especially at low $Q^2$,
since the DGLAP fits to conventional gluon density
are avaibale only for $Q^2 \gsim 1$ GeV$^2$ and can differ
significantly from the $k_t$-factorization resuls, 
see \cite{IN2001}. It would be still interesting to see, 
whether the experimental data possess such a scaling.

\begin{figure}[!htb]
   \centering
   \epsfig{file=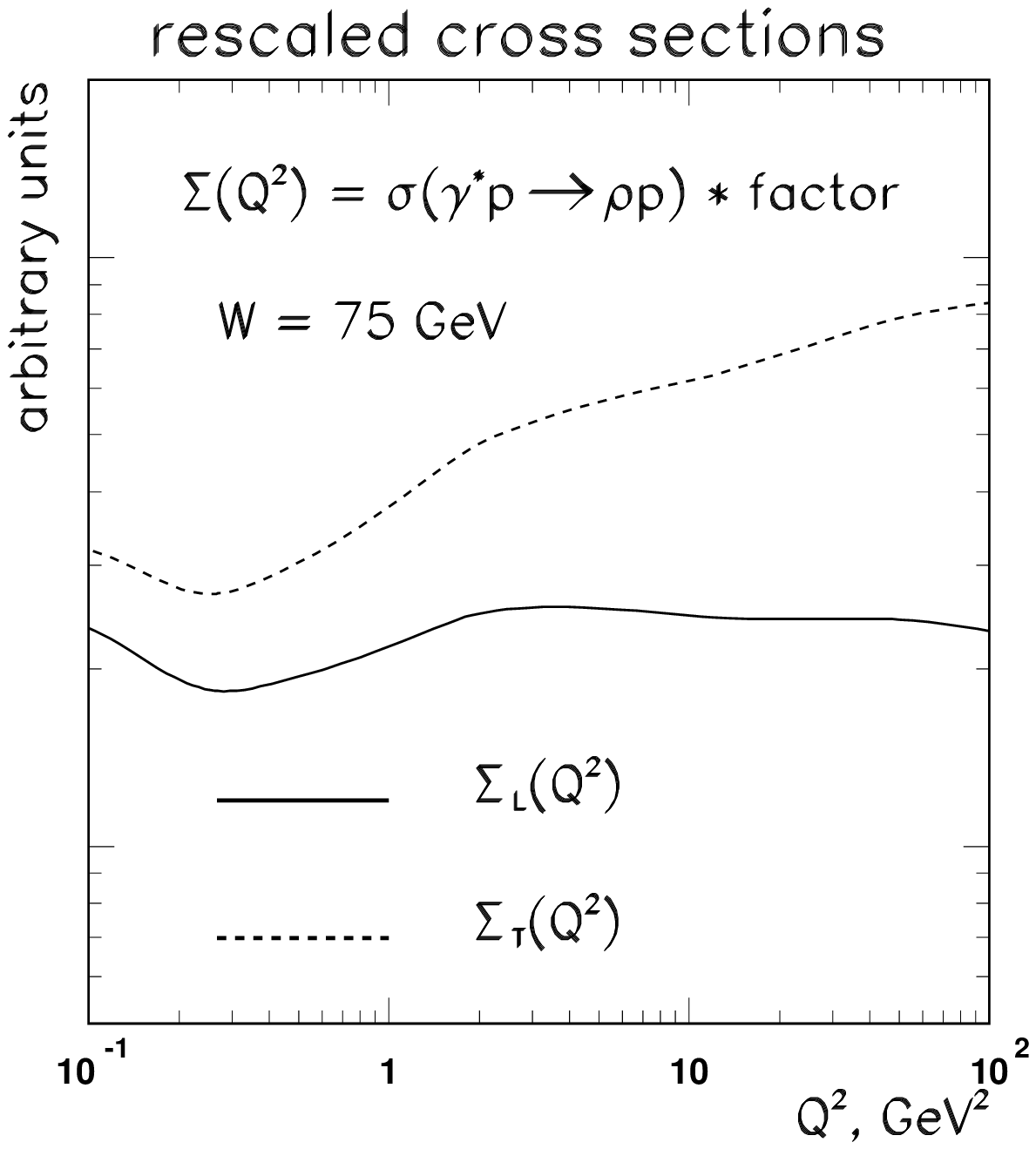,width=8cm}
   \epsfig{file=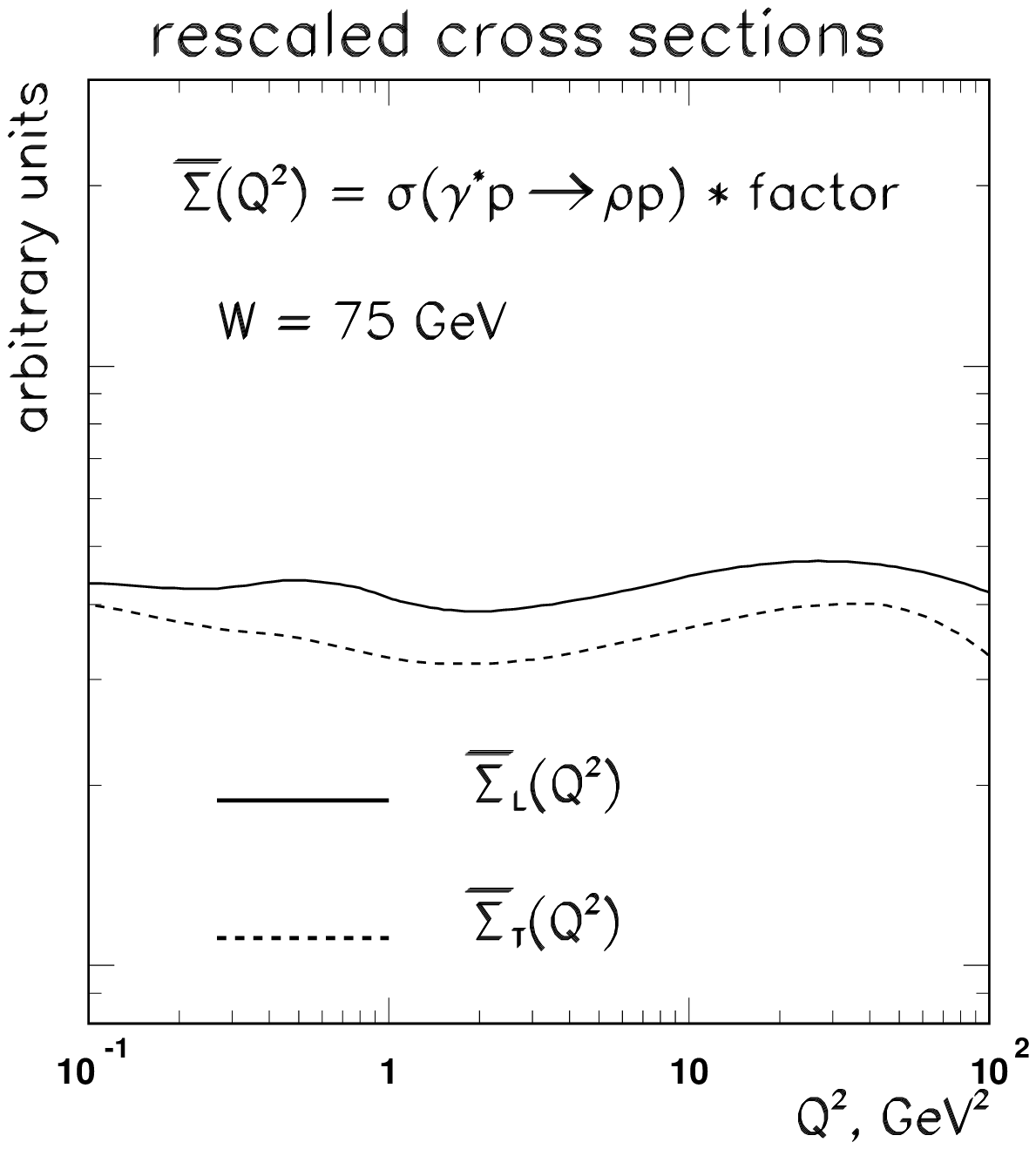,width=8cm}
   \caption{\em (left panel) The rescaled cross sections $\Sigma_L(Q^2)$ 
and $\Sigma_T(Q^2)$ defined according to (\ref{rescaled2}). 
The $Q^2$ behavior of $\Sigma_L(Q^2)$ and $\Sigma_T(Q^2)$ are very different.
(right panel) The rescaled cross sections $\tilde \Sigma_L(Q^2)$ 
and $\tilde \Sigma_T(Q^2)$ defined according to (\ref{rescaled3}). 
The two quantities are close to each other and both have flat $Q^2$ behavior.}
   \label{figrescaled2}
\end{figure}

In the light of the results of the previous section,
it is somewhat surprising that (\ref{rescaled1}), inspired
by the heavy meson limit, is nevertheless rather flat.
If, however, we take a look at $\sigma_L$ and $\sigma_T$ separately,
and define
\be
\Sigma_L(Q^2) = {\sigma_L \over Q^2}\cdot 
{\Qb^8 \cdot b(Q^2)\over \left[G\left(x_g,\Qb^2 \right)
\cdot \alpha_s(\Qb^2)\right]^2}\,;\quad
\Sigma_T(Q^2) = {\sigma_T \over m_\rho^2}\cdot 
{\Qb^8 \cdot b(Q^2)\over \left[G\left(x_g,\Qb^2 \right)
\cdot \alpha_s(\Qb^2)\right]^2}\,,
\label{rescaled2}
\ee
we will see that this remarkable $Q^2$-independence holds
for $\Sigma_L(Q^2)$, but not for $\Sigma_T(Q^2)$, Fig.~\ref{figrescaled2},
left panel.
This difference between $\Sigma_L(Q^2)$ and $\Sigma_T(Q^2)$
can be regarded as a {\em direct proof} that the hard
scales relevant for longitudinal and transverse cross sections
are different.

If we take into account the results of the previous section
and introduce
\be
\tilde \Sigma_L(Q^2) = {\sigma_L \over Q^2}\cdot 
{(\varepsilon_L^2)^4 \cdot b(Q^2)\over \left[G\left(x_g,\Qb_L^2 \right)
\cdot \alpha_s(\varepsilon_L^2)\right]^2}\,;\quad
\tilde \Sigma_T(Q^2) = {\sigma_T \over m_\rho^2} \cdot 
{(\varepsilon_T^2)^4 \cdot b(Q^2)\over \left[G\left(x_g,\Qb_T^2 \right)
\cdot \alpha_s(\varepsilon_T^2)\right]^2}\,,
\label{rescaled3}
\ee
then these rescaled quantities have flat $Q^2$ shape both
for longitudinal and transverse cross sections,
Fig.\ref{figrescaled2}, right panel. Moreover, $\tilde \Sigma_L(Q^2)$ and 
$\tilde \Sigma_T(Q^2)$ are close to each other,
which should be expected by the construction of the ``effective twists"
and factorization scales. These curves are not exact constants
due to non-zero average value of $\vec\Delta^2$ and
due to the presence of helicity-flip amplitudes.

Again, it would be very interesting to see a similar analysis
of the experimental data in terms of $\tilde \Sigma_i(Q^2)$. 
In particular, it would be interesting to see whether
such an analysis can help resolve the long-standing
$\sigma_T$-problem of the $k_t$-factorization predictions 
\cite{VMtalks,disser}.

\section{Discussion and conclusions}

The most of the analysis of the prevous sections
was done with the oscillator wave function. 
The same analysis with other wave function Ans\"atze
will yield slightly different numbers in (\ref{qb2param}) and (\ref{vareps}).
Nevertheless, the general picture will remain the same.
Namely, several observations are stable against the variations 
of the exact shape of the wave function:
\begin{itemize}
\item at smaller $Q^2$ ($Q^2 \lsim 3\div 5$ GeV$^2$)
the DGLAP factorization scale is larger than $(Q^2 + m_\rho^2)/4$;
\item
at large enough $Q^2$ ($Q^2 \gsim 10$ GeV$^2$ for the transverse 
amplitude and $Q^2 \gsim 20$ GeV$^2$ for the longitudinal amplitude), 
the factorization scale is significantly smaller than $(Q^2 + m_\rho^2)/4$.
This should be taken as a word of caution against an unwarranted
application of the DGLAP approach to the problem
of $\rho$ meson production even at high $Q^2$;
\item
the overall $Q^2$ dependence of the pQCD factorization scale
is significantly flatter than $(Q^2 + m_\rho^2)/4$. This is
mostly due to the specific way the $\vec\kappa^2$-behavior of the 
unintegrated gluon density changes, as the $Q^2$ increases
(at fixed $W$). 
\item
the presence of $m_\rho$ in the often used scale $(Q^2+m_\rho^2)/4$
is misleading, since the $\rho$ meson mass has little relevance to
the color dipole interaction with the target proton.
Instead, $Q^2$ appears in combinations of the form of $Q^2 + M^2$
with $M^2 \approx 1 \div 1.5$ GeV$^2$.
\end{itemize}

The numerically accurate understanding of the relevant
hard scales in the $\rho$ production allowed us to
study in detail the $Q^2$ dependence of the $\rho$ 
production cross section. With the help of the rescaled cross sections
(\ref{rescaled1})--(\ref{rescaled3}), Figs.~\ref{figrescaled1} and \ref{figrescaled2},
we showed once more that the production of transverse and longitudinal 
vector mesons is governed by distinct hard scales.
When we took into account the difference of the scales just
found, we observed a very close similarity between the rescaled 
cross sections $\tilde \Sigma_L(Q^2)$ and $\tilde \Sigma_T(Q^2)$.
It would be very interesing to see the results of a similar analysis
of the experimental data. This analysis might help resolve
the long-standing problem of too small $\sigma_T$ at high $Q^2$.

Let us also comment on a contribution to the recent puzzle of
energy independence of $r_\rho = \sigma(\gamma^{*} p \to \rho p)/ 
\sigma_{tot}(\gamma^{*}p)$ ratio. As was discussed in this paper,
at higher $Q^2$, the true factorization scale is smaller 
than $(Q^2+m_\rho^2)/4$. Inverting the argument, one can state
that $\sigma_{tot}(Q_{tot}^2 = (Q^2 + m_\rho^2)/4)$
is expected to be harder than $\sigma_\rho(Q^2)$.
This should enhance the expected value of the Pomeron
intercept in $\sigma_{tot}$ and reduce
the prediction of the energy behavior of $r_\rho$.
Unfortunately, our numerical analysis showed that this effect
is marginal and does not lead to resolution of the problem.\\

I am thankful to Kolya Nikolaev for many valuable comments
and to Igor Akushevich for his help in the early
stage of the code development.
I also wish to thank Prof.~J.Speth for hospitality 
at the Institut f\"ur Kernphysik, Forschungszentrum J\"ulich.
The work was supported by INTAS grants 00-00679 and 00-00366,
and RFBR grant 02-02-17884, and grant ``Universities of Russia" 
UR 02.01.005.

\end{document}